\documentclass[conference]{IEEEtran}
\IEEEoverridecommandlockouts

\usepackage{cite}
\usepackage{amsmath,amssymb,amsfonts}
\usepackage{algorithmic}
\usepackage{graphicx}
\usepackage{textcomp}
\usepackage{xcolor}
\usepackage{enumerate}

\usepackage[final]{changes} %final
\usepackage{subcaption}
\usepackage[bookmarks=false]{hyperref}
\usepackage{balance}
\def\BibTeX{{\rm B\kern-.05em{\sc i\kern-.025em b}\kern-.08em
    T\kern-.1667em\lower.7ex\hbox{E}\kern-.125emX}}

\title{Towards Anatomy Education with Generative AI-based Virtual Assistants in Immersive Virtual Reality Environments
}

\let\oldtwocolumn\twocolumn
\renewcommand\twocolumn[1][]{%
    \oldtwocolumn[{#1}{
    \begin{center}
        \includegraphics[width=\textwidth]{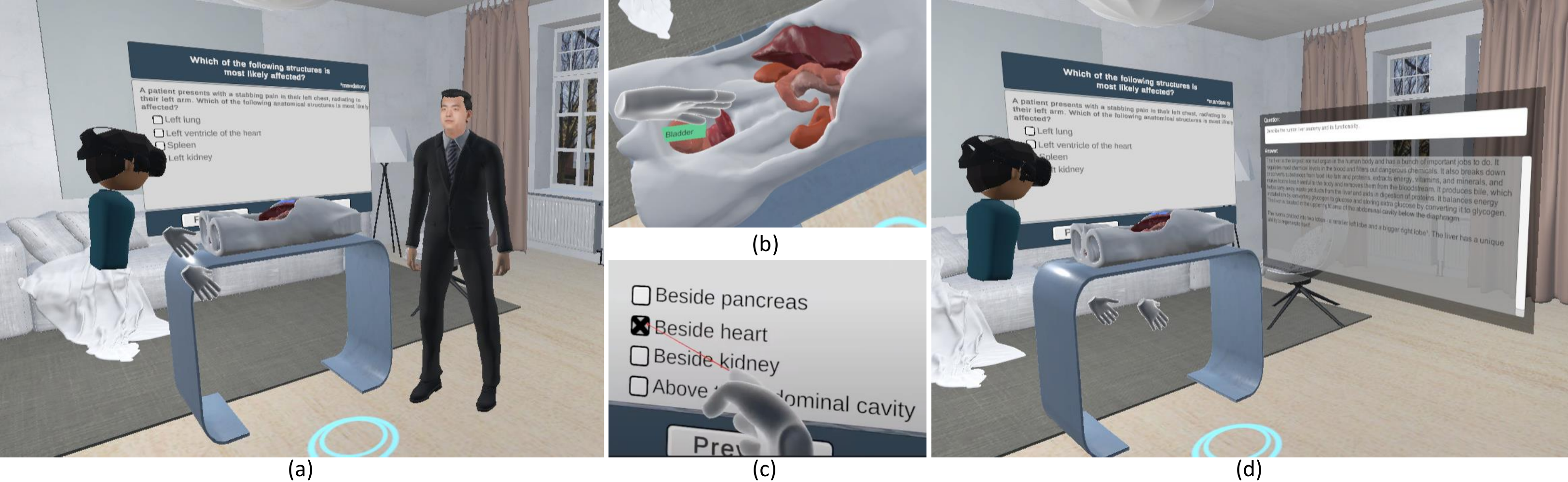}
        \captionof{figure}{\added{An immersive VR environment for human anatomy education with generative AI virtual assistants: (a) the user interacts with the virtual assistant as an avatar representation, (b) interactions with abdominal organ 3D models, (c) the user interacts with UI to answer quiz, and (d) configurations of a screen-based virtual assistant.}}
           \label{fig:teaser}
        \end{center}
    }]
}

\begin{document}
%\IEEEoverridecommandlockouts

%% Author
\author{\IEEEauthorblockN{Vuthea Chheang$^{1}$$^{*}$\thanks{$^{*}$The study was conducted while Vuthea Chheang was a postdoctoral researcher at the University of Delaware. Email: chheang1@llnl.gov.}, %0000-0001-5999-4968
Shayla Sharmin$^{2}$, %0000-0001-5137-1301
Rommy M\'{a}rquez-Hern\'{a}ndez$^{2}$, %0000-0002-7124-2467
Megha Patel$^{2}$, %0009-0000-6156-6161
%Shayla Sharmin$^{2}$, %0000-0001-5137-1301
Danush Rajasekaran$^{2}$, %0009-0005-2785-3446
\\Gavin Caulfield$^{2}$, %0009-0009-3219-6996\\
Behdokht Kiafar$^{2}$, %0009-0001-4415-1332
Jicheng Li$^{2}$, %0000-0003-2564-6337
Pinar Kullu$^{2}$, %
Roghayeh Leila Barmaki$^{2}$$^{\S}$\thanks{$^{\S}$Email: rlb@udel.edu.}%0000-0002-7570-5270 %rlbl@udel.edu
}
\IEEEauthorblockA{$^{1}$Center for Applied Scientific Computing, Lawrence Livermore National Laboratory, Livermore, CA, United States}
\IEEEauthorblockA{$^{2}$Department of Computer and Information Sciences, University of Delaware, Newark, DE, United States
}

}
%\thanks{$^{*}$ The study was conducted while Vuthea Chheang was a postdoctoral researcher at the University of Delaware. \\
%$^{\S}$ Corresponding to Roghayeh Leila Barmaki (rlb@udel.edu).}

\maketitle

\begin{abstract}

Virtual reality (VR) and interactive 3D visualization systems have enhanced educational experiences and environments, particularly in complicated subjects such as anatomy education. VR-based systems surpass the potential limitations of traditional training approaches in facilitating interactive engagement among students. 
However, research on embodied virtual assistants that leverage generative artificial intelligence (AI) and verbal communication in the anatomy education context is underrepresented.
In this work, we introduce a VR environment with a generative AI-embodied virtual assistant to support participants in responding to varying cognitive complexity anatomy questions and enable verbal communication.
We assessed the technical efficacy and usability of the proposed environment in a pilot user study with 16 participants.
We conducted a within-subject design for virtual assistant configuration (\textit{avatar-} and \textit{screen-based}), with two levels of cognitive complexity (\textit{knowledge-} and \textit{analysis-based}). 
The results reveal a significant difference in the scores obtained from \textit{knowledge-} and \textit{analysis-based} questions in relation to \textit{avatar} configuration. %Consequently, participants engaged in a more significant way with the \textit{avatar} as compared to the \textit{screen}-based configuration. 
Moreover, results provide insights into usability, cognitive task load, and the sense of presence in the proposed virtual assistant configurations.  
Our environment and results of the pilot study offer potential benefits and future research directions beyond medical education, using generative AI and embodied virtual agents as customized virtual conversational assistants.
\end{abstract}

% \begin{figure*}[t]
%   \includegraphics[width=\textwidth]{Figures/Overview.pdf}
%   \caption{An immersive VR environment for human anatomy education with generative AI virtual assistant: (a) the user interacts with the virtual assistants as an avatar representation, (b) interactions with abdominal organ 3D models, (c) the user interacts with UI to answer quiz, and (d) the configuration of a screen-based virtual assistant.}
%   \label{fig:environment}
% \end{figure*}

\section{Introduction} 
\noindent
Medical anatomy education, an essential aspect of medical training, necessitates learning the structures and functions of the anatomy in the human body.
These skills are vital prerequisites for surgical procedures. Therefore, student awareness of the variation in morphology and the locations of anatomical structures hold significance. Traditionally, medical students learn human anatomy through textbooks, lectures, and dissection of cadavers. However, these approaches have several limitations, such as lack of interactivity, cost, and ethical considerations of cadaveric dissections~\cite{PREIM2018132}. 
Traditional methods of assessing anatomy knowledge encompass a range of approaches, including spotter, written, and oral examinations ~\cite{smith2015integrated}. For example, anatomy education and assessment have been enhanced by Anderson's modified Bloom's taxonomy, namely Bloom-Anderson principles~\cite{anderson2001taxonomy}. The adoption of this taxonomy for anatomy education has a twofold function: first, it considers the cognitive complexity of assignment questions; second, it provides valuable feedback to learners in the context of anatomy education~\cite{smith2015integrated, morton2017measuring}. 

In recent years, VR has emerged as a viable alternative to traditional anatomy education approaches \cite{chen2017recent, chheang2021collaborative, preim2021virtual}. VR enables students to immerse themselves in an engaging and interactive virtual environment where they can interact with 3D anatomy models. %$Virtual and augmented reality help students reach the top of the Bloom-Anderson learning paradigm \cite{anderson2001taxonomy} with increased effectiveness of interactive learning \cite{Nori2022}. %The use of VR simulators for written and oral Bloom taxonomy evaluations demonstrated significant changes in higher-order cognitive skills % 
%The application of VR simulators in assessments based on Bloom's taxonomy, both written and oral, has shown notable enhancements in higher-order cognitive skills \cite{rupasinghe2011virtual}.
In addition, VR allows medical students to conveniently learn via virtual training without worrying about ethical reservations \cite{de2016virtual, chheang2020toward}. %Moreover, VR allows users to take advantage of virtual forums, gamification, peer learning, and virtual laboratories to foster collaborative and learner-centered environments that align well with Bloom's taxonomy 
VR also empowers learners to leverage virtual forums, gamification, peer learning, and virtual laboratories, fostering collaborative and learner-centered environments that align seamlessly with the principles of Bloom's taxonomy \cite{Nori2022}.
%One of the major challenges is the lack of personalized learning experiences. 

However, most VR-based systems for anatomy education rely on pre-programmed, fixed scenarios that may not adapt well to meet individual learning needs. 
Here, the support of generative artificial intelligence (AI), such as ChatGPT \cite{sallam2023chatgpt}, can be exceptionally advantageous.
%This is where generative artificial intelligent (AI) assistance, e.g., ChatGPT, can be beneficial..
Compared to conventional virtual assistants, which can be rigid based on predefined rules and templates, generative AI technologies, including chatbots and embodied virtual assistants, have the ability to generate more natural and engaging dialogues that resemble human-human interactions. Although chatbots previously relied on pattern matching and string processing \cite{lo2023impact, chen2023artificial}, current chatbots use AI, natural language processing (NLP), large-language models (LLMs), and knowledge-based algorithms \cite{fitria2023chatbots}. These novel technologies empower chatbots to provide more accurate, in-context, personalized, and swift responses to user input while replicating human-human conversations and adapting to the context, levels, and interests of users \cite{hsu2023termbot, tam2023nursing}. 
Moreover, generative AI leverages large-scale data and information from various sources, including scientific articles, textbooks, and datasets. It generates rich and diverse content to enhance communication and understanding.

%In this work, we investigate the effects of cognitive complexity in an anatomical context within a virtual reality (VR) environment. 
This work presents an immersive VR environment designed to support human anatomy education using generative AI conversational assistance (see~\autoref{fig:teaser}).
We integrated generative AI services (ChatGPT-3.5, OpenAI) into the VR environment, enhanced the embodied virtual avatar representation and realism with lip synchronization, and proposed a new framework to tackle the conversational communication between the user and the virtual assistant using several connected services.
The proposed environment has the potential to offer students a more interactive, adaptive, and informative learning experience by offering an embodied virtual assistant. %/companion/tutor. 
To assess the feasibility of the proposed environment, we developed two different configurations of interaction (\textit{avatar}- and \textit{screen}-based virtual assistants) with two levels of cognitive complexity (\textit{knowledge-} and \textit{analysis-}based) and compared them in a within-subjects pilot user study (n\,=\,16).
Our contributions are as follows:
\begin{itemize}
    \setlength\itemsep{0em}
    \item An immersive VR environment to support the human anatomy education, enabling users to communicate verbally and interact with generative AI-based embodied virtual assistants.
    
    \item Results of a pilot user study (n\,=\,16) that provides insights into user performance, usability, task load, and sense of presence in the VR environment.

    \item An exploratory analysis aimed at identifying potential features, benefits, limitations, and research directions.
\end{itemize}

%We organized the remainder of the paper as follows. In the next section, we list the related work. In Section \ref{method}, we provide details on the participants, the apparatus, the study procedure, and the study design. In Section \ref{results}, we report the study results. In Section \ref{discussion}, we discuss the findings, limitations, and implications. Finally, we conclude this paper in Section \ref{conclusion}.

\section{Related Work}
\label{related-work} \noindent
This section provides an overview of previous research on the general use of chatbots for education, with specific focus on VR and virtual assistants in the anatomy education context.

\subsection{Chatbots}
\label{chatbots}
\noindent
Chatbots represent sophisticated computer programs that emulate human-like conversations. They adeptly analyze user inputs and formulate contextually appropriate responses in natural language text. 
They serve as digital platforms that facilitate concurrent interactions between humans and computers. 
Chatbots are widespread in many applications, including e-Commerce, education, healthcare, and entertainment~\cite{caldarini2022literature}. 
The development of chatbots has become increasingly accessible and versatile.
Chatbots such as BERT~\cite{kenton2019bert}, GPT~\cite{brown2020GPT}, and Llama~\cite{touvron2023llama} have pioneered advancements in NLP, while newer iterations %like Transformer-XL~\cite{dai2019transformerXL} 
extend their ability in context understanding. 
These chatbots can assist in the classroom by addressing uncertainties, promoting learning, and providing medical education materials~\cite{hsu2023termbot}. 

Instructors can benefit from their help with scheduling, student concerns, and technological support~\cite{tam2023nursing}. 
Chatbots also provide flexible learning help at the convenience of learners, regardless of time or location~\cite{fitria2023chatbots}. 
In the context of medical education, Termbot~\cite{hsu2023termbot} offers a convenient way for students to practice medical terminologies. 
For nursing, AI chatbots are helpful in courses to practice communication, as well as evaluation and intervention skills with patients~\cite{tam2023nursing}. 
Besides, ChatGPT can enable physicians to quickly generate discharge summaries by entering specific facts, concepts, and suggestions~\cite{patel2023chatgpt}. 
Specifically, ChatGPT has recently been tailored to develop a medical safety LLM framework~\cite{howard2023chatgpt}. This framework involved evaluating ChatGPT's antimicrobial advice across eight hypothetical infection scenarios assessing its suitability, consistency, safety, and stewardship.
Finally, within the realm of VR, generative AI with virtual assistants has been used in different applications, including offering support and guidance to individuals with neurological disorders and their caregivers~\cite{cheng2023exploring}, and as an assistive tool for spinal cord surgeries \cite{he2023will}. 

%Through tasks such as answering queries, dispensing information, and suggesting coping mechanisms, GPT-4 has the potential to enhance the overall quality of life for those affected by such conditions. 
% Additionally, \cite{he2023will} underscores the potential of ChatGPT/GPT-4 as a valuable tool for spinal surgeons \cite{he2023will}, provided it is used judiciously and responsibly. 
% \cite{cheng2023potential} claimed that AI-powered virtual assistants could significantly benefit surgeons specializing in joint arthroplasty, ultimately leading to improved therapeutic outcomes in joint replacement procedures and heightened patient satisfaction.

%%%%%%%%%%%%%%%%%%%%%%%%%%%%%%%%%%%%%%%%%%%%%%%%%%%%%%%%%%%%%%%%%%%%%%%%%%%%
%%%%% More citations
\subsection{VR-based Anatomy Education Systems}
\label{VR}

\noindent VR and augmented reality (AR) technologies have shown potential in improving medical anatomy education 
\cite{pedram2023toward, makinen2022user,Juan2022Learning,barmaki2019enhancement, bork2017exploring}. 
Kurt et al. \cite{KURT2013109} discuss various medical anatomy training approaches, highlighting that cadaver training is restrained by model availability, ethical concerns, and health risks. As an alternative, VR-based training that simulates real-life events sounds appealing.
VR-based training can reduce risks, training time, and cost while engaging students.
%Traditional methods of learning medical anatomy can be enhanced by using different VR and augmented reality (AR) systems to provide a superior learning experience \cite{pedram2023toward, makinen2022user, bracq2021training, Juan2022Learning}. 
%Kurt et al. \cite{KURT2013109} discuss different training methods available for medical anatomy. 
%The training approach using cadavers has significant limitations due to the difficulty of obtaining the model, ethical issues, and the health risks involved. VR-based training replicating real-life situations, on the other hand, has gained popularity as an alternative approach. The potential advantages of VR-based training include minimizing risks, reducing training time and cost, and allowing students to learn in an engaging environment.

A survey by Preim \cite{PREIM2018132} emphasizes the importance of learning perspectives, compelling scenarios, and encouraging strategies in anatomy education.
%A survey conducted by \cite{PREIM2018132} highlighted the importance of focusing on learning perspectives and emphasizing the importance of designing compelling learning scenarios and motivational strategies to support anatomy education.
Erolin et al. \cite{erolin2019using} introduced 3D anatomical models for VR anatomy instruction. In their pilot study, most participants found the models easy to interact with and gave positive feedback. 
Nakai et al. \cite{nakai2022anatomy} explored the potential of VR-based anatomy courses covering nervous, musculoskeletal, and cardiovascular systems for medical students. They created a VR environment that allows users to manipulate organ anatomy. The findings suggested that the study might provide many 3D models and real-time collaboration. 
%Erolin et al.\cite{erolin2019using} presented a collection of 3D anatomical models to enhance anatomy education in VR. Their pilot results indicate that most participants found the models easy to interact with and expressed positive feedback. 
%Nakai et al. \cite{nakai2022anatomy} discussed the feasibility of hosting anatomy lectures for medical students that involve the nervous, musculoskeletal, and cardiovascular systems in VR format. They developed the VR environment, allowing participants to interact and modify the anatomical structures of internal body organs. The results suggested potential advantages of the study, including access to multiple 3D models and real-time collaboration. 

Kurul et al.\cite{kurul2020alternative} studied the impact of immersive VR on anatomy training in physical therapy.
Their findings emphasize the value of VR as an alternative training tool. Falah et al.\cite{6918271} created a VR environment to teach students about medical procedures and anatomical structures. Their solution lets users modify medical data into 3D representations and adjust object sizes in the virtual world.
%Kurul et al.\cite{kurul2020alternative} conducted a study investigating how immersive VR affects anatomy training in undergraduate physical therapy students. Their results reveal a difference between the VR and control groups and highlight the importance of using VR as an alternative method of anatomy training. Falah et al.\cite{6918271} developed a VR environment to train students to recognize and comprehend anatomical structures and medical procedures. Their system allows users to manipulate medical data, transforming it into 3D representations and adjusting the relative sizes of objects within the virtual environment.
Izard et al. \cite{Izard2016} found that VR effectively improves anatomical comprehension in a similar study employing the cranium anatomy.
To enhance medical students' learning experience, Saalfield et al.\cite{Saalfeld_2020} developed a tutoring system that allows teachers to assist students in learning human skull anatomy in a shared virtual environment. 
Schott et al. \cite{schott2021vr} proposed a multi-user VR/AR environment for liver anatomy education utilizing clinical examples.
% A multi-user VR/AR environment for liver anatomy education utilizing clinical examples was also proposed by Schott et al. \cite{schott2021vr}. 
According to the study, the prototype could help surgery education in small learning groups and classrooms.

 Overall, most of these systems used 3D anatomical models for information visualizations in VR/AR, but they lacked the use of AI or LLMs to make these experiences more learner-centered or self-paced. Our work leverages the power of VR for information visualization by leveraging generative AI and LLMs to offer more genuine interactions with learners.

%In a similar approach, Izard et al. \cite{Izard2016} conducted a study using cranium anatomy and yielded similar results, demonstrating the efficacy of VR in enhancing anatomical understanding. Saalfield et al.\cite{Saalfeld_2020} introduced a tutoring system to improve medical students' learning experience in a shared virtual environment; thus, a teacher can help the student to learn the anatomy of the human skull. Schott et al. \cite{schott2021vr} proposed a multi-user VR / AR environment for liver anatomy education using clinical cases. The results demonstrate the potential benefits of the prototype to facilitate surgery education and support a range of training scenarios, from small learning groups to classroom-size settings.

%In addition to VR systems, AR also shows the potential to support anatomy learning environments and could serve as a supplementary teaching tool~\cite{bork2021effectiveness, minopoulos2023medical, richards2023student}. 
%Results show that students' behavioral engagement was enhanced in student autonomy and time allocation towards task completion, specifically with viewing images and answering related questions.
\subsection{Embodied Virtual Assistants}
\label{VR Embodiment}
\noindent Anatomy education in VR environments entails a complex and resource-intensive process. 
It may require a suitable environment and competent instructors to teach human anatomy effectively.
The psychological implications of virtual assistants have been a focus of active research \cite{moro2017effectiveness, mao2021immersive, bernardo2017virtual, norouzi2018systematic, barmaki2018embodiment,yu2018accuracy}.
Kim et al.~\cite{8613756} conducted a study investigating different levels of embodiment in virtual assistants controlled by human-in-the-loop.% They created an avatar with various animation movements and a human-in-the-loop with three different conditions: \textit{speech}, \textit{speech and gesturing}, and \textit{speech, gesturing, and locomotion}. 
The study suggested that gesturing and locomotion of the avatar increased trust between the user and the interactive virtual assistant. 
Haesler et al. \cite{8699187} conducted a study using \textit{Amazon Alexa}, comparing a voice-only version with an \textit{embodied avatar} version to perform simple everyday tasks.
The results showed that the participants preferred the embodied avatar over the voice-only version. In a separate study, Kim et al. \cite{9089596} analyzed the reduction of task loads with \textit{Amazon Alexa} %in three conditions: the user working alone, a voice-only assistant, and an embodied assistant. 
and findings indicated that using a voice assistant reduced the number of tasks; however, users still expressed uncertainty regarding tasks outside their visual range. % being completed by the voice-only assistant.
On the other hand, the embodied version instilled more confidence in users regarding task completion, thereby fostering greater trust and collaboration between the assistant and the user \cite{yao2022virtual}.
% summary
These studies collectively indicate that users tend to place more trust in tasks they can visually confirm, highlighting the importance of visual representation in virtual assistants.

% comparison
%Compared to previous research listed in Sections \ref{chatbots} to \ref{VR Embodiment}, 
Compared to previous work, our VR environment offers a unique advantage by integrating a generative AI-based virtual assistant, \textit{ChatGPT, OpenAI}, as a companion to support learning human anatomy. 
The embodied virtual assistant enables users to engage in verbal conversations and receive responses to their information queries, resulting in greater confidence and participation.
Moreover, it can be used as a source of guidance and to provide users with detailed information. Therefore, the user can seek clarification, ask follow-up questions, and receive personalized explanations to facilitate the understanding of complex medical knowledge as a replacement for a human-in-the-loop assistant or a human trainer.

\section{Materials and Methods} \label{method}

\noindent The research questions (RQs) to guide our work are the following:
\begin{enumerate}[\indent {}]\item\textbf{RQ1} How do \textit{configurations} of \textit{avatar}- and \textit{screen}-based virtual assistants influence user performance in anatomy education?

\item\textbf{RQ2} To what extent do subjective measures, such as usability, task load, and presence, associate with virtual assistant configurations?

\item\textbf{RQ3} What are the advantages, limitations, and potential research directions of using generative AI for anatomy education?
\end{enumerate}

\noindent In the following sections, we describe participants, apparatus, study procedure, and study design.  Specifically, \autoref{fig:teaser} to \autoref{fig:studyprocedure} elaborate more about our pilot user study.
\begin{figure*}[!t]
    \centering
    \includegraphics[width=\textwidth]{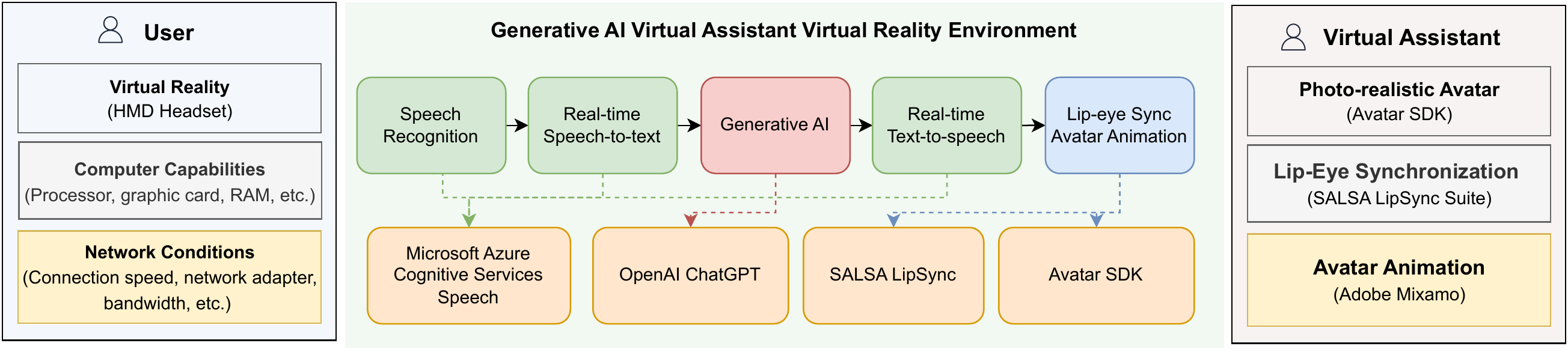}
    \caption{System architecture of our proposed VR environment with the generative AI-based embodied virtual assistant. }
    \label{fig:systemarchitecture}
\end{figure*}

\subsection{Participants}
%%\textcolor{red}{-- Rommy }

\noindent We conducted a priori power analysis to evaluate the sample size with \textit{analysis of variance} (ANOVA) for interaction effects (repeated measures, within factors). Utilizing the \textit{G*Power} statistical analysis software, we calculated the effect size $\eta_{p}^{2}= 0.14$ for two factors, resulting in a sample size of 16~\cite{faul2007g}.

%During participant recruitment, we implemented a two-step pre-screening process to ensure the selection of suitable candidates. %The first phase was included in the pre-questionnaire that participants had to fill out before scheduling a time to participate in the study. 
%All prospective participants were first required to complete a screening questionnaire prior to scheduling a meeting.
%This initial phase aimed to exclude individuals who self-reported vision or motion sickness concerns that could hinder their interaction in the VR environment, as well as those below the age of 18.  
%The second pre-screening phase was integrated into the training process before the study began. During this phase, participants were assessed for their ability to effectively interact with the VR environment and their comfort level with VR technology.

The study was approved by the University of Delaware's Institution Review Board (\#$2136140-1$). The inclusion criteria were participants over the age of 18 with normal to corrected vision, with no known prior history of motion or cyber sickness. Out of the 20 registrants, four participants were excluded due to vision (one individual), motion sickness (two individuals), and VR discomfort (one individual). 
Thus, 16 participants from the University of Delaware were successfully recruited in our study. The demographic information of the participants are provided in ~\autoref{tab:participants}.

\begin{table}[t]
%\centering
\caption{Participant background and characteristics ($n=16$).}
\label{tab:participants}
% \scriptsize
\resizebox{\columnwidth}{!}{
\begin{tabular}{p{0.42\columnwidth}p{0.12\columnwidth}p{0.15\columnwidth}} %p{0.3\columnwidth}p{0.1\columnwidth}p{0.1\columnwidth}} %p{0.5\columnwidth}p{0.2\columnwidth}p{0.15\columnwidth}
\hline
Characteristics  & Value  & Mean \\
\hline
Age   & [20 -- 35] & 26\,$\pm$\,4.97 \\
Gender                    &    &    \\
\hspace*{0.3cm} Male      & 6 & (37.50\%) \\
\hspace*{0.3cm} Female    & 10 & (62.50\%) \\
Education                &   &      \\
\hspace*{0.3cm} Bachelor's program  & 6 & (37.50\%) \\
\hspace*{0.3cm} Master's program & 3 & (18.75\%) \\
\hspace*{0.3cm} Doctoral program & 7 & (43.75\%) \\
% Video game experience    &   &      \\
% \hspace*{0.3cm}None      & 4 & (25.00\%) \\
% \hspace*{0.3cm}Several times a year & 7 & (43.75\%) \\
% \hspace*{0.3cm}Several times a week & 3 & (18.75\%) \\
% \hspace*{0.3cm}Daily     & 2 & (12.50\%) \\
Medical anatomy knowledge           &    &     \\
\hspace*{0.3cm}Not much   & 5 & (12.50\%) \\
\hspace*{0.3cm}Basic & 11 & (87.50\%) \\
Health centered classes           &    &     \\
\hspace*{0.3cm}No   & 8 & (50.00\%) \\
\hspace*{0.3cm}Yes  & 8 & (50.00\%) \\
VR experience            &    &     \\
\hspace*{0.3cm}Never used before   & 6 & (37.50\%) \\
\hspace*{0.3cm}Used a few times & 3 & (18.75\%) \\
\hspace*{0.3cm}Used several times   & 5 & (31.25\%) \\
\hspace*{0.3cm}Regular use   & 2 & (12.50\%) \\
Handedness           &    &     \\
\hspace*{0.3cm}Left   & 2 & (12.50\%) \\
\hspace*{0.3cm}Right   & 14 & (87.50\%) \\

\hline
\end{tabular}
}
\end{table}

\subsection{Apparatus}
%%\textcolor{red}{-- Vuthea}
\noindent
\autoref{fig:systemarchitecture} shows an overview of the system architecture for the proposed VR environment designed for human anatomy education, featuring generative AI virtual assistants.
The VR environment was developed using \textit{Unity} game engine (version $2019.4.34f1$). 
Customization of 3D models, including the living room, and the incorporation of additional models from \textit{Sketchfab}, alongside \textit{OpenHELP} organ models, were integral to the VR environment's development \cite{kenngott2015openhelp}.
Participants were asked to navigate the VR system and engage with the 3D model by grasping, resizing, and rotating them to understand their functionalities during the training session. %The 3D model of the body displayed on the table remained constant throughout the training session.
%The 3D models of the living room and additional models from \textit{Sketchfab} were customized, and \textit{OpenHELP} organ models were used in the VR environment \cite{kenngott2015openhelp}. 
We used the \textit{Valve Index} VR headset, controllers, lighthouses, and its components within the VR setup.

To enhance user interactions within the VR environment, we utilized the capabilities of the \textit{Virtual Reality Toolkit} (VRTK). This toolkit enabled us to implement fundamental interactions such as teleportation, object manipulation, and interactions with the user interface (UI). Moreover, we used the VR questionnaire toolkit~\cite{feick2020virtual} to develop the UI for the VR quiz.
% \textit{Virtual Reality Toolkit} (VRTK) was used to develop basic interactions, including teleportation, grabbing objects, and interactions with user interface (UI).  
% The UI for the VR quiz was developed with the VR questionnaire toolkit~\cite{feick2020virtual}. 
%To provide users with an intelligent conversational agent for answering questions, we integrated \textit{ChatGPT, OpenAI, USA}.
We integrated \textit{ChatGPT} (OpenAI, USA) to provide services as an intelligent conversational agent for answering questions.
We also used an AI-based library (\textit{Avatar SDK}, Itseez3D Inc., USA) to create a photo-realistic model for the virtual assistant's avatar presented in the user study. 
%Furthermore, our VR environment featured a virtual avatar generated using the AI-based library Avatar SDK by Itseez3D Inc., USA, resulting in a photorealistic model.
% We built the virtual avatar through an AI-based library (Avatar SDK, Itseez3D Inc., USA) in order to simulate a photorealistic model.
This virtual avatar was animated to provide gestures with facial expressions, further enhancing user engagement. We leveraged the Microsoft Azure Speech service to enable natural interactions, utilizing text-to-embodiment capabilities for text-to-speech and speech-to-text. %Text-to-embodiment refers to the conversion of the participant's speech to text and subsequently into the avatar's voice and mouth expressions.
Text-to-embodiment refers to the conversion of text responses from generative AI into the virtual avatar's voice and facial expressions and vice versa, e.g., participant's speech to text.

\begin{figure*}
    \centering
    \includegraphics[width=\textwidth]{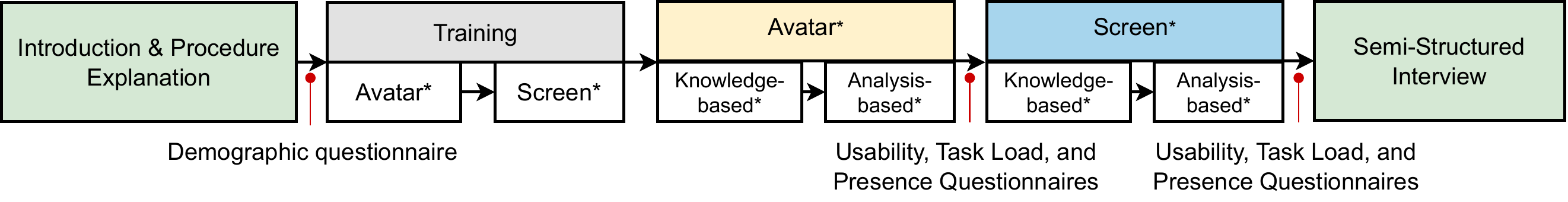}
    \caption{Overview of the study procedure. The order of the conditions (marked with $^*$) was counterbalanced.}
    \label{fig:studyprocedure}
    
\end{figure*}

\subsection{Study Procedure}
\noindent \autoref{fig:studyprocedure} presents an overview of the study procedure.
To accommodate potential learning curves, we counterbalanced the order of the conditions and the level of cognitive complexity. We walked each participant through the study's procedure. Ensuring their full understanding of the consent form was a priority before requesting their signature.
The participants were given the opportunity to become familiarized with the VR headset and controllers and learn how to interact with the virtual assistant in the training environment. 
They received guidance on asking questions to the virtual assistant by pressing a button on the controller while posing their query, then releasing it to allow the system to process and respond. Additionally, they learned the distinctions in responses between the avatar and the text screen.
% Once the participants felt confident using the technology, we began the study. 
The study commenced upon participants' confidence on the technology.

% \textit{\textbf{Tasks}}

To start the study, participants found themselves in a virtual living room environment with a virtual cadaver, a screen displaying a question with multiple-choice answers, and their initial virtual assistant.
Interacting within this environment, they could teleport around the room, move the screens around, and manipulate the organs within the cadaver. When ready, participants selected the ``Next'' button to receive their first question. They were informed they could ask the virtual assistant as many questions as they wished to arrive at the answer to the multiple-choice question, with no pressure to provide a correct response. 
The only condition was that they could not ask the virtual assistant the entire question directly.

After obtaining an answer, participants selected it from the multiple-choice options and clicked the ``Submit'' button. Subsequently, they were presented with another question featuring a different cognitive level within the same virtual assistant configuration.
Upon completing tasks with their first virtual assistant configuration, participants took a break and completed a short mid-questionnaire on a computer. Following this, they repeated the process of answering two questions with varying levels of cognitive complexity using the other virtual assistant configuration. Once this phase concluded, participants were again given a break to complete a post-questionnaire. 
Lastly, they had the opportunity to ask any questions and provide qualitative feedback.

% \begin{figure}[h]
%     \centering
%     \includegraphics[width=\columnwidth]{Figures/Avatar}
%     \caption{Environment with avatar.}
%     \label{fig:avatar}
% \end{figure}

% \begin{figure}[h]
%     \centering
%     \includegraphics[width=\columnwidth]{Figures/Screen}
%     \caption{Environment with screen.}
%     \label{fig:screen}
% \end{figure}

\subsection{Study Design}
\noindent
Our study was designed as a $2\times2$  within-subject experiment ($2$ \textit{configurations} $\times$ $2$ \textit{levels of cognitive complexity}).  
Each participant was randomly assigned to start with either the \textit{avatar} or the \textit{screen} configuration during the study. They were then presented with two questions of varying cognitive complexity levels for each virtual assistant configuration. We ensured that the opportunity to initiate with either the avatar or the screen was evenly distributed among all participants. This random assignment and counterbalancing helped mitigate potential bias resulting from a learning effect.

% The opportunity to start with either the \textit{avatar} or the \textit{screen} was counterbalanced between all participants. 
% This was to avoid bias due to the learning effect and to ensure an equal number of participants performing each order. 

\subsubsection{\textbf{Independent Variables}}

In our study, we defined the virtual assistant configuration and difficulty level as independent variables.

\paragraph{\textbf{Configuration}} Within the virtual environment, we introduced two types of generative AI virtual assistants: \textit{avatar} and \textit{screen} (see also ~\autoref{fig:teaser}). 

\begin{itemize}
    \setlength\itemsep{0em}
    \item \textit{Avatar}: an embodied avatar equipped with audio and lip synchronization, seamlessly integrated with \textit{Microsoft Azure} text-to-speech and generative AI services to respond to questions.
    
    % \item \textit{Screen}: a screen displayed text responses generated by generative AI services, accompanied by the participant's question. 

    \item \textit{Screen}: a screen displayed text responses generated by generative AI service alongside the participant's question.
    
\end{itemize}

\paragraph{\textbf{Level of Cognitive Complexity}}
%\paragraph{Level of Cognitive Complexity}
We had four questions categorized in two sets. %of two that were aligned with Bloom's taxonomy cognitive complexity levels.
\begin{itemize}

\item  \textit{Knowledge-based}: These questions required no analytical thinking or in-depth understanding and fell under the foundational or first level of Bloom's taxonomy, namely ``knowledge" or ``remembering". 

\item \textit{Analysis-based}: These questions demanded more in-depth analysis, corresponding to the fourth level of Bloom's taxonomy, namely ``analysis".

% We have four questions based on Bloom's Taxonomy cognitive complexity level, categorized into two sets \textit{Knowledge-based} and \textit{Analysis-based}. 
%In the first set, two questions fell under the foundational level of ``knowledge" or ``remembering" in Bloom's taxonomy. 

%In the first set, the questions required no analytical thinking or in-depth understanding. The other questions in \textit{Analysis-based} presented scenarios that demand analysis, corresponding to the fourth level of Bloom's taxonomy, namely ``analyzing" or ``analysis".

%There are two levels of difficulty: \textit{easy} and \textit{hard}. 

%We listed our questions (Qs) below. 
%Our question sets (Q1 and Q2) were categorized as \textit{Knowledge-based}, while Q3 and Q4 were identified as \textit{Analysis-based}.
\end{itemize}

These four multiple-choice, scenario-based questions focus on diagnosing medical conditions based on specific symptoms and involved anatomical structures. 
It includes how different symptoms, such as chest pain, difficulty breathing, abdominal pain, and neurological problems, can be linked to specific organs or systems of the human body, such as the heart, lungs, digestive organs, and nervous system. The goal is to determine the most likely affected area or organ responsible for the presented symptoms.
%Each participant received a randomized sequence of cognitively challenging questions. 
Each configuration contains one knowledge-based and one analysis-based questions. The configuration and the order of questions were counterbalanced.

\subsubsection{\textbf{Dependent Variables}}

Throughout the study, we recorded the participant's selected answer, task completion time, and the number of interactions with the virtual assistant as the dependent variables. All this data was automatically logged into a \textit{CSV} file for further analysis. %The study took about 30 to 40 minutes per participant. 

\begin{itemize}
    \setlength\itemsep{0em}
    \item \textit{Task Completion Time}: 
    % the duration from when the participant initiated the question until they submitted the answer. 
    the duration between the participant posing the question and submitting the answer.
    
    \item \textit{Number of Interactions}: the number of times the participant requests information from the virtual assistant.
    
    \item \textit{Score}: a variable indicating whether their answer to the question was correct or incorrect.
    
    %A binary variable that shows whether the participant's answer to the multiple choice question was correct or incorrect. On multiple-choice questions, three responses were correct and one was correct. Participants scored 1 for each right response to the four multiple-choice questions. 
    % A binary variable indicates if the participant answered the multiple-choice question correctly, where three options were incorrect and one was correct. Each correct answer to the four multiple-choice questions earned 1 point. 
    %a binary variable indicating whether the participant answered each of the question correctly. For each correct selection from the set of four choices, where three are incorrect and one is correct, one point was earned.
\end{itemize}

\begin{table*}[!t]
\centering
\caption{Summary of descriptive results of user performance.}
\label{tab:descriptiveResults}
\begin{tabular}{lrrr}
\hline
Variable & Task Completion Time (s) & Number of Interactions & Scores \\
\hline
Avatar & 159.23 (149.25) [26.38] & 5.96 (11.87) [2.10] & 0.59 (0.49) [0.08]  \\
\hspace*{0.2cm} Knowledge-based & 117.74 (72.61) [18.15] & 5.31 (11.22) [2.80] & 0.75 (0.44) [0.11] \\
\hspace*{0.2cm} Analysis-based & 
200.72 (192.59) [48.14] & 6.62 (12.83) [3.20] & 0.43 (0.51) [0.12]\\
Screen & 159.06 (152.94) [27.03] & 3.65 (2.74) [0.48] & 0.50 (0.50) [0.09] \\
\hspace*{0.2cm} Knowledge-based & 177.50 (200.57) [50.14] & 4.31 (3.36) [0.84] & 0.56 (0.51) [0.12] \\
\hspace*{0.2cm} Analysis-based & 140.62 (85.96) [21.49] & 3.00 (1.82) [0.45] & 0.43 (0.51) [0.12] \\
\hline
\end{tabular} 
\\
\raggedright
\hspace*{3.5cm}
\textit{All entities are in the following format: mean value (standard deviation) [standard error].}
\end{table*}

\begin{figure*}[!t]
    \centering     \includegraphics[width=\textwidth]{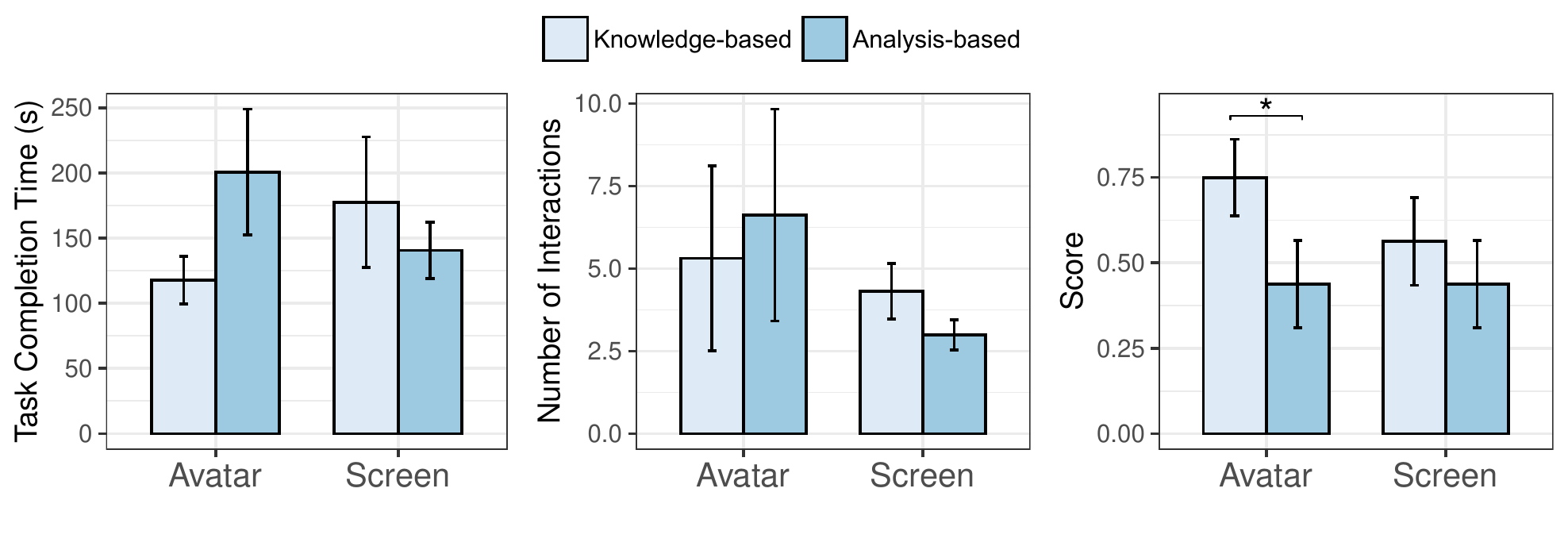}
    \caption{Results of user performance (n\,=\,16), including (left) task completion time, (middle) number of interactions with the virtual assistant, and (right) score. %* denotes significance.
    }
    \label{fig:userperformance}    
\end{figure*}

\subsubsection{\textbf{Questionnaires}}

As part of our evaluation, we gathered not only performance data but also valuable insights into participants' subjective experiences through the administration of standardized questionnaires. These questionnaires were designed and administered using the \textit{Qualtrics} survey platform.

The following are the specific dimensions we assessed:

\begin{itemize}
    \setlength\itemsep{0em}
    \item \textit{Usability}: We assessed usability using the System Usability Scale (SUS) questionnaire \cite{Brooke1995SUS}, which comprises ten questions with a 5-point Likert-scale from ``\textit{strongly disagree}'' to ``\textit{strongly agree}''.
    The final SUS score was calculated on a scale from 0 to 100 (0\,--\,50: not acceptable, 51\,--\,67: poor, 68: okay, 69\,--\,80: good, 81\,--\,100: excellent) \cite{bangor2009determining}. 

    \item \textit{Task Load}: To gauge the subjective task load experienced by participants, we employed the NASA Task Load Index (NASA TLX) questionnaire \cite{hart2006nasa}. This questionnaire consists of six questions assessing mental demand, physical demand, temporal demand, performance, effort, and frustration. 
    
    \item \textit{Presence}: We also evaluated the sense of presence within the virtual environment using  \textit{igroup Presence Questionnaire} (IPQ) \cite{schubert2001experience, schwind2019using}. This questionnaire has 14 questions categorized as general presence, spatial presence, involvement, and experienced realism. Responses are recorded on a 7-point Likert scale, ranging from ``\textit{strongly disagree}'' to ``\textit{strongly agree}''.

\end{itemize}

\subsubsection{\textbf{Semi-structured Interviews}}

After the completion of all previous tasks, we solicited qualitative feedback from participants through semi-structured interviews. Participants were asked the following questions:

\begin{itemize}
    \setlength\itemsep{0em}
    \item What is your feedback on the VR environment and configurations of the virtual assistant?
    \item Do you have any questions or suggestions?
\end{itemize}

% \subsubsection{Data Analysis}
% %%\textcolor{red}{-- Megha \& Rommy \& Vuthea}

% \textit{RStudio} with R for statistical computing was used for data analysis.
% A two-way analysis of variance (ANOVA) was conducted for dependent variables.

% \begin{table}[!t]
% \centering
% \caption{Summary of statistical results ($p<.05$).}
% \label{tab:anovaresults}
% \resizebox{\columnwidth}{!}{
% \begin{tabular}{lllrrcr}
% \hline
% Variable  & DFn & DFd  & F & p & Sig. & $\eta_{p}^{2}$  \\
% \hline
% Task Completion Time & & & & & & \\
% \hspace*{0.2cm} Configuration & 1 & 15 & 0.0001 & 0.992 &  & \textless 0.0001%3.42$e^{-07}$ 
% \\
% \hspace*{0.2cm} Difficulty & 1 & 15 & 3.41 & 0.08 &  & 0.01 \\
% \hspace*{0.2cm} Configuration * Difficulty & 1 & 15 & 3.24 & 0.09 &  & 0.04 \\
% Number of Interactions & & & & & & \\
% \hspace*{0.2cm} Configuration & 1 & 15 & 0.75 & 0.40 &  & 0.02 \\
% \hspace*{0.2cm} Difficulty & 1 & 15 & 0.00 & 1.00 &  & 0.00 \\
% \hspace*{0.2cm} Configuration * Difficulty & 1 & 15 & 4.472 & 0.05 & *  & 0.006 \\
% Score & & & & & & \\
% \hspace*{0.2cm} Configuration & 1 & 15 & 1.00 & 0.33 &  & 0.01 \\
% \hspace*{0.2cm} Difficulty & 1 & 15 & 4.62 & 0.046 & * & 0.073 \\
% \hspace*{0.2cm} Configuration * Difficulty & 1 & 15 & 0.41 & 0.53 &  & 0.01 \\
% \hline
% \end{tabular}
% }
% \end{table}

\section{Results}
\label{results}
% In the following sections, we present the results for both descriptive and statistical analysis of our study. 
% We start by providing detailed descriptions and explanations of user performance, offering a clear view of how participants performed during the study. 
% Then, we give  the outcomes of the questionnaires, evaluating the responses and insights gathered from our participants. 
% Finally, we share the general feedback we received, giving  a complete picture of what our participants had to say. 
% Our goal is to ensure a comprehensive understanding of our findings, both in terms of quantitative data and the insights gained from participants' perspectives. %This thorough examination will enhance understanding of the significance and implications of our research.
\noindent In the following sections, we present the results for both descriptive and statistical analysis of user performance, questionnaire outcomes, and general feedback.

\subsection{User Performance Results (RQ1)}
\noindent We used \textit{RStudio} with the R programming language to conduct a thorough statistical analysis. 
Our selected method, a two-way ANOVA for dependent variables, enabled us to examine the variables in depth. 
As we explore our analysis further, a summary of descriptive results related to objective user performance measures is shown in both tabular (\autoref{tab:descriptiveResults}) and graphical (\autoref{fig:userperformance}) formats. 
% We used \textit{RStudio} with R for computing the statistical analysis with a two-way ANOVA for dependent variables.
% A summary of descriptive results for objective user performance measures is shown in Table~\ref{tab:descriptiveResults} and Figure~\ref{fig:userperformance}.
%The statistical analysis results are listed in \ref{tab:anovaresults}.  

\subsubsection{Task Completion Time (TCT)}
We found no significant differences between the \textit{configuration}, \textit{level of cognitive complexity} and their interaction effect on task completion time. 
On average, participants responded faster to \textit{knowledge-based} questions in the \textit{avatar} configuration. In the \textit{screen} configuration, \textit{knowledge-based} questions were responded more slowly. %than\textit{analysis-based}.  
%Conversely, in the \textit{screen} configuration, \textit{knowledge-based} questions took longer to answer than \textit{analysis-based} ones.
The results show that participants spent more time interacting with the \textit{avatar} when tackling \textit{analysis-based} questions, while the \textit{screen} configuration led to more extended task durations for \textit{knowledge-based} questions.
Descriptive results for total completion time, however, showed only a minimal difference between the \textit{avatar} (\textit{M} = 159.23 s, \textit{SD} = 149.25) and \textit{screen} (\textit{M} = 159.06 s, \textit{SD} = 152.94) configurations. 
It indicates that both configurations are comparable to assist the user in solving the tasks.
% We found no significant differences between the \textit{configuration}, \textit{level of cognitive complexity} and their interaction effect on task completion time. 
% %For \textit{level of difficulty}, \textit{Easy} questions were on average faster than \textit{Hard} in the \textit{avatar} configuration, while in the \textit{screen} configuration, they were vice versa.
% Answering the \textcolor{green}{\textit{knowledge-based}} questions were on average faster than \textit{analysis-based} in the \textit{avatar}
% configuration. While in the \textit{screen} configuration, \textit {knowledge based} questions were on average slower than \textit{analysis based} while answering.
% The results showed participants spent more time interacting with the \textit{avatar} in the \textcolor{green}{\textit{analysis-based questions}}, while it took more time to complete tasks on the \textit{screen}, e.g., reading texts, in the \textit{knowledge-based} question level. 
% Descriptive results for total completion time, however, show a minimal difference between \textit{avatar}: 159.23\,s (\textit{SD} = 149.25) and \textit{screen}: 159.06\,s (\textit{SD} = 152.94) configuration.
% It indicates that both configurations are comparable to assist the user in solving the tasks.

\subsubsection{Number of Interactions}

Descriptive results show that all tasks in the \textit{avatar} configuration require more interactions and requests with the virtual assistant, particularly in the context of \textit{analysis-based} questions. 
%However, there is a trend for significant difference \textbf{(\textit{p} = 0.05)} between the \textit{configuration}, \textit{level of cognitive complexity}, and their interaction effect on the number of interactions.
The results could indicate that users need more explanation through verbal communication with the virtual avatar compared to the display screen.%% This observation aligns with previous literature as well, because when it comes to learning new concepts in anatomy, students reported to learn more effectively visually rather than auditorily~\cite{bucsan2014learning}, so learning via virtual avatar narration resulted in more interactions from students. 

\subsubsection{Score}

On average, participants scored higher in solving questions within the \textit{avatar} configuration than in the \textit{screen} configuration.
Regarding the \textit{level of cognitive complexity}, there was a significant difference between configurations (\textit{F}(1, 15) = 4.62, \textbf{\textit{p} = 0.046}, $\eta_{p}^{2}$ = 0.07). 
Subsequent pairwise t-tests indicated a statistically significant difference between \textit{knowledge-based} and \textit{analysis-based} questions (\textit{t} = -1.78, \textit{df} = 15, \textbf{\textit{p} = 0.04}) within the \textit{avatar} configuration (see~\autoref{fig:userperformance}). 
This finding shows that participants who engaged with \textit{knowledge-based} questions in the \textit{avatar} configuration obtained higher scores and completed tasks more swiftly compared to \textit{analysis-based} questions. This finding aligns with the Bloom's taxonomy about the level of cognitive complexity as well, because \textit{knowledge-based} questions are less cognitively demanding.

\begin{figure*}[t]
    \centering     \includegraphics[width=\textwidth]{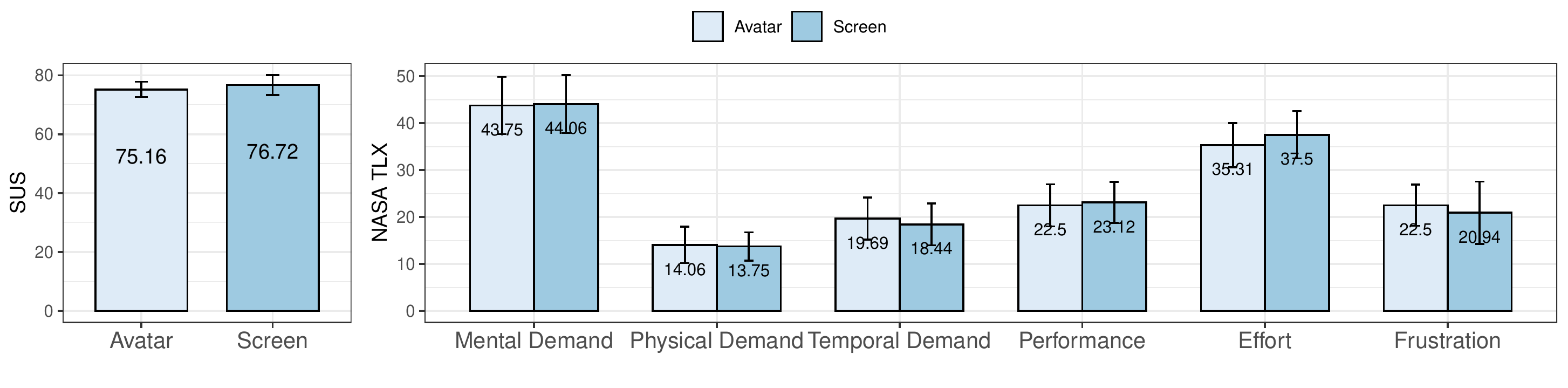}
    \caption{Results of the questionnaires: (left) system usability scale (SUS) and (right) NASA task load index (TLX).}
    \label{fig:sus_tlx}    
\end{figure*}

\begin{figure}[h]
    \centering     \includegraphics[width=\columnwidth]{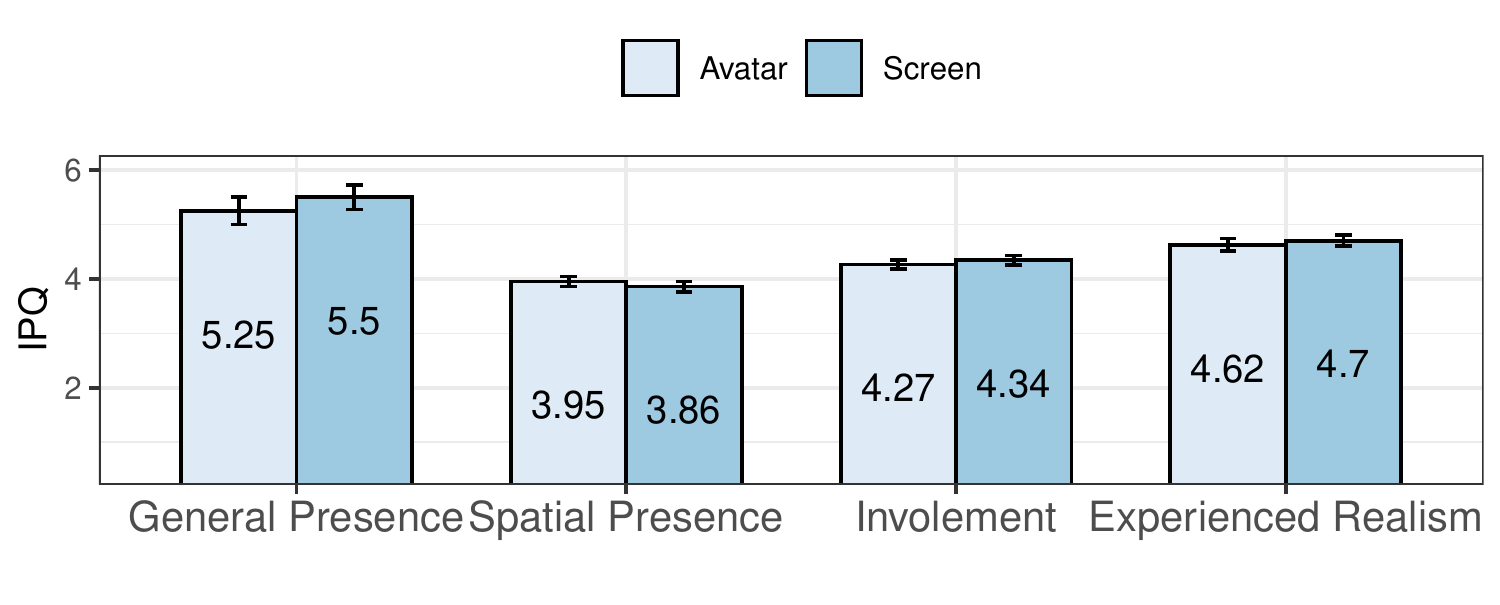}
    \caption{Sense of presence results using \textit{Igroup} presence questionnaire (IPQ) in the virtual environment.}
    \label{fig:ipq}    
\end{figure}

\subsection{Questionnaire Results (RQ2)}
%%\textcolor{red}{-- Megha \& Rommy \& Vuthea}
\noindent
In the following sections, we present the subjective results obtained from our questionnaires, offering insights into usability, task load, and presence within our virtual environment. 

\subsubsection{Usability}
\noindent
The results of the usability assessment using the SUS questionnaire yielded an average score for \textit{avatar} (\textit{M} = 75.16, \textit{SD} = 10.55) and \textit{screen} (\textit{M} = 76.72, \textit{SD} = 13.53) (see~\autoref{fig:sus_tlx}). 
No significant differences were observed between the main effects and their interaction effect.
Both configurations scored higher than 68, indicating above-average usability \cite{bangor2009determining} and highlighting their potential usability benefits.

\subsubsection{Task Load}
\noindent
The subjective task load was evaluated using an unweighted (raw) NASA-TLX questionnaire.  
Descriptive results, as shown in~\autoref{fig:sus_tlx}, demonstrated no significant differences in task load between configurations for all load dimensions (i.e., mental demand, physical demand, temporal demand, performance, effort, and frustration). 
Notably, mental demand (\textit{avatar}: \textit{M} = 43.75, \textit{SD} = 24.32; \textit{screen}: \textit{M} = 44.06, \textit{SD} = 24.71) emerged as the highest-rated load dimension, closely followed by effort (\textit{avatar}: \textit{M} = 35.31, \textit{SD} = 18.83; \textit{screen}: \textit{M} = 37.50, \textit{SD} = 20.00). 
% It can be seen that the highest scoring of them was mental demand: \textit{avatar} (\textit{M} = 43.75, \textit{SD} = 24.32), \textit{screen} (\textit{M} = 44.06, \textit{SD} = 24.71) closely followed by effort: \textit{avatar} (\textit{M} = 35.31, \textit{SD} = 18.83), \textit{screen} (\textit{M} = 37.50, \textit{SD} = 20.00). In contrast, physical demand: \textit{avatar} (\textit{M} = 14.06, \textit{SD} = 15.40), \textit{screen} (\textit{M} = 13.75, \textit{SD} = 12.04) was the lowest scoring.  
% This suggests that the proposed environment may primarily impact mental demand as participants engaged with the virtual assistant to solve the questions.
This implies the proposed environment mainly impacts mental demand as participants engage with the virtual assistant to solve questions.
%%\textcolor{red}{Describe more on its items.}

\subsubsection{Presence}
\noindent
The sense of presence in the immersive VR environment was assessed using an IPQ questionnaire. 
Notably, no significant differences were observed among the configurations. 
Descriptive results indicated average scores for general presence (\textit{avatar}: \textit{M} = 5.25, \textit{SD} = 1.00; \textit{screen}: \textit{M} = 5.50, \textit{SD} = 0.89), spatial presence (\textit{avatar}: \textit{M} = 3.95, \textit{SD} = 0.36; \textit{screen}: \textit{M} = 3.86, \textit{SD} = 0.37), involvement (\textit{avatar}: \textit{M} = 4.27, \textit{SD} = 0.32; \textit{screen}: \textit{M} = 4.34, \textit{SD} = 0.35), and experienced realism (\textit{avatar}: \textit{M} = 4.62, \textit{SD} = 0.46; \textit{screen}: \textit{M} = 4.70, \textit{SD} = 0.41). 
Descriptively, general presence and experienced realism garnered the highest scores, followed by involvement and spatial presence (see~\autoref{fig:ipq}).

% The sense of presence in the immersive VR environment was assessed using an IPQ questionnaire. 
% There were no significant differences among the configurations. Descriptive results show the average score for general presence: \textit{avatar} (\textit{M} = 5.25, \textit{SD} = 1.00), \textit{screen} (\textit{M} = 5.50, \textit{SD} = 0.89), 
% spatial presence: \textit{avatar} (\textit{M} = 3.95, \textit{SD} = 0.36), \textit{screen} (\textit{M} = 3.86, \textit{SD} = 0.37), 
% involvement: \textit{avatar} (\textit{M} = 4.27, \textit{SD} = 0.32), \textit{screen} (\textit{M} = 4.34, \textit{SD} = 0.35) , 
% and experienced realism: \textit{avatar} (\textit{M} = 4.62, \textit{SD} = 0.46), \textit{screen} (\textit{M} = 4.70, \textit{SD} = 0.41)). 
% For descriptive results, it could show that the general presence and experienced realism had the highest score, followed by involvement and spatial presence (see Figure~\ref{fig:ipq}).

\subsection{Qualitative Participant Feedback (RQ3)}

\noindent
Participants offered open-ended feedback expressing their experiences about the difference between the two configurations of the study. Two participants preferred the \textit{avatar} for heightened immersion. But there were suggestions regarding the improvement of voice synchronization in \textit{avatar} condition. Participants also noted the need to pre-plan their questions, as delayed inquiries sometimes led to conversation interruptions by the system. Additionally, some participants occasionally forgot to release the talk button, causing the virtual assistant's on-screen response to be overridden by new text reacting to their continued speech.

\section{Discussion}
\label{discussion}
\noindent

The proposed VR environment was evaluated in a pilot study to assess the feasibility and usability of a generative AI-based question-answering prototype with embodied and non-embodied virtual assistants. 

%\paragraph{Effect of Virtual Assistant Configurations (RQ1)}

\textit{RQ1} focused on the impact of the two virtual assistant configurations on user performance.
% concerned how the two virtual assistant configurations would impact user performance. 
We found that participants scored significantly higher when answering \textit{knowledge-based} as compared to \textit{analysis-based} questions in the \textit{avatar} configuration.
The fact that we did not find significant differences on \textit{screen} leads us to believe that participants might have had an easier time keeping track of the virtual assistant's answers for the \textit{analysis-based} questions in the \textit{screen} than the \textit{avatar} configuration. 
%\noindent We also observed significant differences between the configuration and the level of cognitive complexity with respect to the number of interactions.

Trends in descriptive results of the number of interactions, the tasks in \textit{avatar} configuration require more interactions with the virtual assistant.
This is supported by the fact that the participants had more interactions with the virtual assistant in the \textit{avatar} than the \textit{screen} with the \textit{analysis-based} questions. Additionally, regarding user feedback, the participants did not show a clear preference for either the \textit{screen} or the \textit{avatar}. It could also indicate that they found benefits in both scenarios for different types of questions. As such, we can gain additional insights by gathering more information on the benefits of the \textit{screen} compared to the \textit{avatar} configuration, specifically considering the levels of cognitive complexity. Knowing what makes each configuration helpful will allow us to combine the two configurations in a way that will be best for user performance in anatomy education.

%\paragraph{Subjective Measures (RQ2)}
\textit{RQ2} pertained to the impact of subjective measures on the virtual assistant configuration. There was no clear significance regarding usability, task load, or presence. Generally, it seemed that the participants were experiencing higher mental demand than physical demand. Additionally, the participants appeared to score higher for general presence than spatial presence.
These preferences could be influenced by various factors, including the study location, experience with virtual assistants and VR, and potential distractions that prevented participants from forgetting their surroundings.
Regarding usability, both configurations could qualify as relatively easy to use, which could add to our belief that each configuration has clear potential benefits. Therefore, combining both configurations could provide a full option for interacting with the environment.
%, and we need to gather more information on the screen scenario's benefits compared to the avatar scenario.

%\paragraph{Potential Benefits, Limitations, and Research Directions (RQ3)}
\textit{RQ3} concerned the limitations and potential research directions of using generative AI for anatomy education. 
Integrating generative AI virtual assistant could adapt to users and provide personalized support \cite{sallam2023chatgpt}. 
It has the potential to offer an engaging, immersive learning experience, enhancing motivation.
% In addition, it could offer an engaging, immersive, and interactive learning experience, enhancing the learning process and motivation.
The generative AI-based virtual assistant can query a vast database of information and provide comprehensive information and resources according to the student's needs. 
However, evaluating the quality and accuracy of responses remains a crucial aspect that requires further investigation.
% the quality and accuracy of the responses could be another topic needed for further evaluation.

%%% Newly added result
The results of this experiment indicate that generative AI appears more effective at providing direct answers, and the participants achieved higher scores on knowledge-based questions when cognitive complexity was low.
However, analysis problems require more cognitive engagement, interpretation, and complex comprehension and may not be solved easily by the generative AI. 
Such questions involve human-like reasoning, complex judgment, and sometimes subjective interpretation. This shows that cognitive complexity, as measured by Bloom's taxonomy, is correlated with generative AI's ability to support learners.
Generative AI may not solve analysis problems easily that demand human-like thinking, complicated judgment, and subjective interpretation, which require higher cognitive involvement, interpretation, and complex comprehension.
%%%%% newly added result end

\paragraph*{Limitations and Future Work} 
Our pilot study was conducted with 16 participants. Although enough for this study design to show sufficient power, it is still relatively small.
During the study, we noticed that the way a participant phrased a question could impact the response they received.
It correlates with other studies showing that virtual assistants could be biased, particularly when asked to find a relationship between items \cite{alkaissi2023artificial}. %bang2023multitask
For AI-generated contents, there is a concern regarding generating somewhat inaccurate or misleading responses, the lack of transparency and unclear information on the data source used, and other related consequences, as noted in previous work \cite{sallam2023chatgpt}.
Hence, future work should investigate the conversations' transcripts and the responses' accuracy.
Regarding the verbal conversation using speech-to-text, some of our participants had problems being understood. A problem arose for participants with an accent and when they spoke too quickly. Consequently, these factors could have affected the participants' correctness scores and experience. 
The varying levels of experience for users using VR or virtual assistant, such as \textit{ChatGPT} was found to be a challenge as well. 
%It was very clear from watching the participants who had used VR before versus the ones who had minimal to no experience that there was an effect on confidence and immersion.
Participants with little experience were less likely to interact with the environment and look around, which impacted how different they felt the two configurations were. For future studies, levels of experience should be considered as a covariate, or a balance of the groups based on their level of prior VR/virtual assistant experiences would be beneficial.
%Additionally, all but two participants treated the interactions like they would a search engine instead of taking advantage of the more conversational dialogue that the virtual assistant can maintain. We believe this is due to their inexperience with the generative AI and could have impacted their immersion and experience.

%\paragraph{Future Work}
For future work, experience with a larger sample size and further development and integration with visual/object input and output, e.g., ChatGPT-4, could provide an extensive learning environment. Furthermore, providing tools to monitor learning progress and assessments may prove advantageous.
Incorporating new modalities \cite{li2023mmasd} and advanced techniques such as visual/acoustic emotion recognition \cite{hartholt2019ubiquitous, canal2022survey, li21twostage}, gaze engagement tracking \cite{guo2023social}, and body gesture analysis~\cite{barmaki2018gesturing, li2022pose} could improve the representation of virtual assistants. %li2022dyadic, barmaki2016gesture, Li2021improving, 
Additionally, it would be interesting to study the learning effects in a collaborative VR environment \cite{scavarelli2021virtual, chheang2022towards}. %, which allows multiple users to join in the same shared virtual environment \cite{scavarelli2021virtual, chheang2022towards, PREIM2023403}.

\section{Conclusion}
\label{conclusion}
\noindent
In this paper, we have developed an immersive VR environment featuring a generative AI-based embodied virtual assistant designed for human anatomy education.
This environment was evaluated in a pilot user study involving 16 participants with no prior knowledge of the subject. 
The evaluation results demonstrated the impact of virtual assistant configurations on user performance. While there were small differences between the \textit{avatar} and \textit{screen}-based configurations in terms of the number of interactions, a significant difference emerged in the cognitive complexity level of questions associated with the \textit{avatar}-based configuration.  
Additionally, we reported subjective measure results from usability, task load, and sense of presence. 
The combination of both virtual assistant configurations has the potential to offer a comprehensive solution for assisting and enhancing the learning experience.
% Combining both virtual assistant configurations could provide a comprehensive option to assist and enhance the learning experience. 
Moreover, our findings provide insights into potential benefits, limitations, and future research directions concerning the utilization of embodied virtual agents and generative AI conversational chatbots in education. 
%Our proposed environment offers interactive, adaptive, and informative learning opportunities that meet training and education needs.

%% identification of the section in the article metadata and the
% \begin{acks}

%%% Do not forget to thank Ishwar Kumar M.A. Sekaran and participants who participated in the user study

% \end{acks}



\section*{Acknowledgment}
\noindent 
We thank Ishwar Kumar M.A. Sekaran for his contributions to study recruitment. We are also thankful to our study subjects for their support and participation. We wish to acknowledge the support from our sponsors, the National Science Foundation (2222661-2222663 and 2321274) and the National Institute of General Medical Sciences of the National Institutes of Health (P20 GM103446-E). 
This work was also prepared by LLNL under Contract DE-AC52-07NA27344.
Any opinions, findings, conclusions, or recommendations expressed in this material are those of the authors and do not necessarily reflect the view of the sponsors.

%\noindent We thank study participants, and Ishwar Kumar M.A. Sekaran for their support and participation in our user study. We also wish to acknowledge the support from our sponsors, National Science Foundation (\#2222661-2222663 and \#$2321274$), National Institute of General Medical Sciences of the National Institutes of Health (\#$P20 GM103446-E$), and US Department of Energy (\#$DE-AC52-07NA27344$). Any opinions, findings, conclusions, or recommendations expressed in this material are those of the authors and do not necessarily reflect the view of the sponsors.

\newpage
\balance
\bibliographystyle{IEEEtran}
\bibliography{Main}

\begin{thebibliography}{10}
\providecommand{\url}[1]{#1}
\csname url@samestyle\endcsname
\providecommand{\newblock}{\relax}
\providecommand{\bibinfo}[2]{#2}
\providecommand{\BIBentrySTDinterwordspacing}{\spaceskip=0pt\relax}
\providecommand{\BIBentryALTinterwordstretchfactor}{4}
\providecommand{\BIBentryALTinterwordspacing}{\spaceskip=\fontdimen2\font plus
\BIBentryALTinterwordstretchfactor\fontdimen3\font minus
  \fontdimen4\font\relax}
\providecommand{\BIBforeignlanguage}[2]{{%
\expandafter\ifx\csname l@#1\endcsname\relax
\typeout{** WARNING: IEEEtran.bst: No hyphenation pattern has been}%
\typeout{** loaded for the language `#1'. Using the pattern for}%
\typeout{** the default language instead.}%
\else
\language=\csname l@#1\endcsname
\fi
#2}}
\providecommand{\BIBdecl}{\relax}
\BIBdecl

\bibitem{PREIM2018132}
B.~Preim and P.~Saalfeld, ``A survey of virtual human anatomy education
  systems,'' \emph{Computers \& Graphics}, vol.~71, pp. 132--153, 2018.

\bibitem{smith2015integrated}
C.~F. Smith and B.~McManus, ``The integrated anatomy practical paper: A robust
  assessment method for anatomy education today,'' \emph{Anatomical Sciences
  Education}, vol.~8, no.~1, pp. 63--73, 2015.

\bibitem{anderson2001taxonomy}
L.~W. Anderson and D.~R. Krathwohl, \emph{A taxonomy for learning, teaching,
  and assessing: A revision of Bloom's taxonomy of educational objectives:
  complete edition}.\hskip 1em plus 0.5em minus 0.4em\relax Addison Wesley
  Longman, Inc., 2001.

\bibitem{morton2017measuring}
D.~A. Morton and J.~M. Colbert-Getz, ``Measuring the impact of the flipped
  anatomy classroom: The importance of categorizing an assessment by bloom's
  taxonomy,'' \emph{Anatomical sciences education}, vol.~10, no.~2, pp.
  170--175, 2017.

\bibitem{chen2017recent}
L.~Chen, T.~W. Day, W.~Tang, and N.~W. John, ``Recent developments and future
  challenges in medical mixed reality,'' in \emph{Proc. of IEEE international
  symposium on mixed and augmented reality (ISMAR)}.\hskip 1em plus 0.5em minus
  0.4em\relax IEEE, 2017, pp. 123--135.

\bibitem{chheang2021collaborative}
V.~Chheang, P.~Saalfeld, F.~Joeres, C.~Boedecker, T.~Huber, F.~Huettl, H.~Lang,
  B.~Preim, and C.~Hansen, ``A collaborative virtual reality environment for
  liver surgery planning,'' \emph{Computers \& Graphics}, vol.~99, pp.
  234--246, 2021.

\bibitem{preim2021virtual}
B.~Preim, P.~Saalfeld, and C.~Hansen, ``Virtual and augmented reality for
  educational anatomy,'' in \emph{Digital Anatomy: Applications of Virtual,
  Mixed and Augmented Reality}.\hskip 1em plus 0.5em minus 0.4em\relax
  Springer, 2021, pp. 299--324.

\bibitem{de2016virtual}
J.~W.~V. De~Faria, M.~J. Teixeira, L.~d. M.~S. J{\'u}nior, J.~P. Otoch, and
  E.~G. Figueiredo, ``Virtual and stereoscopic anatomy: when virtual reality
  meets medical education,'' \emph{Journal of neurosurgery}, vol. 125, no.~5,
  pp. 1105--1111, 2016.

\bibitem{chheang2020toward}
V.~Chheang, V.~Fischer, H.~Buggenhagen, T.~Huber, F.~Huettl, W.~Kneist,
  B.~Preim, P.~Saalfeld, and C.~Hansen, ``Toward interprofessional team
  training for surgeons and anesthesiologists using virtual reality,''
  \emph{International journal of computer assisted radiology and surgery},
  vol.~15, pp. 2109--2118, 2020.

\bibitem{Nori2022}
N.~Barari, M.~RezaeiZadeh, A.~Khorasani, and F.~Alami, ``Designing and
  validating educational standards for e-teaching in virtual learning
  environments (vles), based on revised bloom’s taxonomy,'' \emph{Interactive
  Learning Environments}, vol.~30, no.~9, pp. 1640--1652, 2022.

\bibitem{sallam2023chatgpt}
M.~Sallam, ``Chatgpt utility in healthcare education, research, and practice:
  systematic review on the promising perspectives and valid concerns,''
  \emph{Healthcare}, vol.~11, no.~6, p. 887, 2023.

\bibitem{lo2023impact}
C.~K. Lo, ``What is the impact of chatgpt on education? a rapid review of the
  literature,'' \emph{Education Sciences}, vol.~13, no.~4, p. 410, 2023.

\bibitem{chen2023artificial}
Y.~Chen, S.~Jensen, L.~J. Albert, S.~Gupta, and T.~Lee, ``Artificial
  intelligence (ai) student assistants in the classroom: Designing chatbots to
  support student success,'' \emph{Information Systems Frontiers}, vol.~25,
  no.~1, pp. 161--182, 2023.

\bibitem{fitria2023chatbots}
T.~N. Fitria, N.~E. Simbolon, and A.~Afdaleni, ``Chatbots as online chat
  conversation in the education sector,'' \emph{International Journal of
  Computer and Information System (IJCIS)}, vol.~4, no.~3, pp. 93--104, 2023.

\bibitem{hsu2023termbot}
M.-H. Hsu, T.-M. Chan, and C.-S. Yu, ``Termbot: A chatbot-based crossword game
  for gamified medical terminology learning,'' \emph{International Journal of
  Environmental Research and Public Health}, vol.~20, no.~5, p. 4185, 2023.

\bibitem{tam2023nursing}
W.~Tam, T.~Huynh, A.~Tang, S.~Luong, Y.~Khatri, and W.~Zhou, ``Nursing
  education in the age of artificial intelligence powered chatbots
  (ai-chatbots): Are we ready yet?'' \emph{Nurse Education Today}, vol. 129, p.
  105917, 2023.

\bibitem{caldarini2022literature}
G.~Caldarini, S.~Jaf, and K.~McGarry, ``A literature survey of recent advances
  in chatbots,'' \emph{Information}, vol.~13, no.~1, p.~41, 2022.

\bibitem{kenton2019bert}
J.~D. M.-W.~C. Kenton and L.~K. Toutanova, ``Bert: Pre-training of deep
  bidirectional transformers for language understanding,'' in \emph{Proceedings
  of naacL-HLT}, vol.~1, 2019, p.~2.

\bibitem{brown2020GPT}
T.~Brown, B.~Mann, N.~Ryder, M.~Subbiah, J.~D. Kaplan, P.~Dhariwal,
  A.~Neelakantan, P.~Shyam, G.~Sastry, A.~Askell \emph{et~al.}, ``Language
  models are few-shot learners,'' \emph{Advances in neural information
  processing systems}, vol.~33, pp. 1877--1901, 2020.

\bibitem{touvron2023llama}
H.~Touvron, T.~Lavril, G.~Izacard, X.~Martinet, M.-A. Lachaux, T.~Lacroix,
  B.~Rozi{\`e}re, N.~Goyal, E.~Hambro, F.~Azhar \emph{et~al.}, ``Llama: Open
  and efficient foundation language models,'' \emph{arXiv preprint
  arXiv:2302.13971}, 2023.

\bibitem{patel2023chatgpt}
S.~B. Patel and K.~Lam, ``Chatgpt: the future of discharge summaries?''
  \emph{The Lancet Digital Health}, vol.~5, no.~3, pp. e107--e108, 2023.

\bibitem{howard2023chatgpt}
A.~Howard, W.~Hope, and A.~Gerada, ``Chatgpt and antimicrobial advice: the end
  of the consulting infection doctor?'' \emph{The Lancet Infectious Diseases},
  vol.~23, no.~4, pp. 405--406, 2023.

\bibitem{cheng2023exploring}
K.~Cheng, Q.~Guo, Y.~He, Y.~Lu, S.~Gu, and H.~Wu, ``Exploring the potential of
  gpt-4 in biomedical engineering: the dawn of a new era,'' \emph{Annals of
  Biomedical Engineering}, pp. 1--9, 2023.

\bibitem{he2023will}
Y.~He, H.~Tang, D.~Wang, S.~Gu, G.~Ni, and H.~Wu, ``Will chatgpt/gpt-4 be a
  lighthouse to guide spinal surgeons?'' \emph{Annals of Biomedical
  Engineering}, pp. 1--4, 2023.

\bibitem{pedram2023toward}
S.~Pedram, G.~Kennedy, and S.~Sanzone, ``Toward the validation of vr-hmds for
  medical education: a systematic literature review,'' \emph{Virtual Reality},
  pp. 1--26, 2023.

\bibitem{makinen2022user}
H.~M{\"a}kinen, E.~Haavisto, S.~Havola, and J.-M. Koivisto, ``User experiences
  of virtual reality technologies for healthcare in learning: An integrative
  review,'' \emph{Behaviour \& Information Technology}, vol.~41, no.~1, pp.
  1--17, 2022.

\bibitem{Juan2022Learning}
J.~J. Reyes-Cabrera, J.~M. Santana-Núñez, A.~Trujillo-Pino, M.~Maynar, and
  M.~A. Rodriguez-Florido, ``{Learning Anatomy through Shared Virtual
  Reality},'' in \emph{Eurographics Workshop on Visual Computing for Biology
  and Medicine}.\hskip 1em plus 0.5em minus 0.4em\relax The Eurographics
  Association, 2022.

\bibitem{barmaki2019enhancement}
R.~Barmaki, K.~Yu, R.~Pearlman, R.~Shingles, F.~Bork, G.~M. Osgood, and
  N.~Navab, ``Enhancement of anatomical education using augmented reality: An
  empirical study of body painting,'' \emph{Anatomical sciences education},
  vol.~12, no.~6, pp. 599--609, 2019.

\bibitem{bork2017exploring}
F.~Bork, R.~Barmaki, U.~Eck, P.~Fallavolita, B.~Fuerst, and N.~Navab,
  ``Exploring non-reversing magic mirrors for screen-based augmented reality
  systems,'' in \emph{2017 IEEE virtual reality (VR)}.\hskip 1em plus 0.5em
  minus 0.4em\relax IEEE, 2017, pp. 373--374.

\bibitem{KURT2013109}
E.~Kurt, S.~E. Yurdakul, and A.~Ata{\c{c}}, ``An overview of the technologies
  used for anatomy education in terms of medical history,'' \emph{Procedia -
  Social and Behavioral Sciences}, vol. 103, pp. 109--115, 2013, international
  Educational Technology Conference.

\bibitem{erolin2019using}
C.~Erolin, L.~Reid, and S.~McDougall, ``Using virtual reality to complement and
  enhance anatomy education,'' \emph{Journal of visual communication in
  medicine}, vol.~42, no.~3, pp. 93--101, 2019.

\bibitem{nakai2022anatomy}
K.~Nakai, S.~Terada, A.~Takahara, D.~Hage, R.~S. Tubbs, and J.~Iwanaga,
  ``Anatomy education for medical students in a virtual reality workspace: A
  pilot study,'' \emph{Clinical Anatomy}, vol.~35, no.~1, pp. 40--44, 2022.

\bibitem{kurul2020alternative}
R.~Kurul, M.~N. {\"O}g{\"u}n, A.~Neriman~Narin, {\c{S}}.~Avci, and B.~Yazgan,
  ``An alternative method for anatomy training: Immersive virtual reality,''
  \emph{Anatomical Sciences Education}, vol.~13, no.~5, pp. 648--656, 2020.

\bibitem{6918271}
J.~Falah, S.~Khan, T.~Alfalah, S.~F.~M. Alfalah, W.~Chan, D.~K. Harrison, and
  V.~Charissis, ``Virtual reality medical training system for anatomy
  education,'' in \emph{Science and Information Conference}, 2014, pp.
  752--758.

\bibitem{Izard2016}
S.~G. Izard and J.~A.~J. M{\'e}ndez, ``Virtual reality medical training
  system,'' in \emph{Proceedings of the fourth international conference on
  technological ecosystems for enhancing multiculturality}, 2016, pp. 479--485.

\bibitem{Saalfeld_2020}
P.~Saalfeld, A.~Schmeier, W.~D`Hanis, H.-J. Rothkötter, and B.~Preim,
  ``{Student and Teacher Meet in a Shared Virtual Reality: A~one-on-one
  Tutoring System for Anatomy Education},'' in \emph{{Eurographics Workshop on
  Visual Computing for Biology and Medicine (EG VCBM)}}, 2020, pp. 55--59.

\bibitem{schott2021vr}
D.~Schott, P.~Saalfeld, G.~Schmidt, F.~Joeres, C.~Boedecker, F.~Huettl,
  H.~Lang, T.~Huber, B.~Preim, and C.~Hansen, ``A vr/ar environment for
  multi-user liver anatomy education,'' in \emph{2021 IEEE Virtual Reality and
  3D User Interfaces (VR)}.\hskip 1em plus 0.5em minus 0.4em\relax IEEE, 2021,
  pp. 296--305.

\bibitem{moro2017effectiveness}
C.~Moro, Z.~{\v{S}}tromberga, A.~Raikos, and A.~Stirling, ``The effectiveness
  of virtual and augmented reality in health sciences and medical anatomy,''
  \emph{Anatomical sciences education}, vol.~10, no.~6, pp. 549--559, 2017.

\bibitem{mao2021immersive}
R.~Q. Mao, L.~Lan, J.~Kay, R.~Lohre, O.~R. Ayeni, D.~P. Goel \emph{et~al.},
  ``Immersive virtual reality for surgical training: a systematic review,''
  \emph{Journal of Surgical Research}, vol. 268, pp. 40--58, 2021.

\bibitem{bernardo2017virtual}
A.~Bernardo, ``Virtual reality and simulation in neurosurgical training,''
  \emph{World neurosurgery}, vol. 106, pp. 1015--1029, 2017.

\bibitem{norouzi2018systematic}
N.~Norouzi, K.~Kim, J.~Hochreiter, M.~Lee, S.~Daher, G.~Bruder, and G.~Welch,
  ``A systematic survey of 15 years of user studies published in the
  intelligent virtual agents conference,'' in \emph{Proceedings of the 18th
  international conference on intelligent virtual agents}, 2018, pp. 17--22.

\bibitem{barmaki2018embodiment}
R.~Barmaki and C.~E. Hughes, ``Embodiment analytics of practicing teachers in a
  virtual immersive environment,'' \emph{Journal of Computer Assisted
  Learning}, vol.~34, no.~4, pp. 387--396, 2018.

\bibitem{yu2018accuracy}
K.~Yu, R.~Barmaki, M.~Unberath, A.~Mears, J.~Brey, T.~H. Chung, and N.~Navab,
  ``On the accuracy of low-cost motion capture systems for range of motion
  measurements,'' in \emph{Medical Imaging 2018: Imaging Informatics for
  Healthcare, Research, and Applications}, vol. 10579.\hskip 1em plus 0.5em
  minus 0.4em\relax SPIE, 2018, pp. 90--95.

\bibitem{8613756}
K.~Kim, L.~Boelling, S.~Haesler, J.~Bailenson, G.~Bruder, and G.~F. Welch,
  ``Does a digital assistant need a body? the influence of visual embodiment
  and social behavior on the perception of intelligent virtual agents in ar,''
  in \emph{IEEE International Symposium on Mixed and Augmented Reality
  (ISMAR)}, 2018, pp. 105--114.

\bibitem{8699187}
S.~Haesler, K.~Kim, G.~Bruder, and G.~Welch, ``Seeing is believing: Improving
  the perceived trust in visually embodied alexa in augmented reality,'' in
  \emph{2018 IEEE International Symposium on Mixed and Augmented Reality
  Adjunct (ISMAR-Adjunct)}, 2018, pp. 204--205.

\bibitem{9089596}
K.~Kim, C.~M. de~Melo, N.~Norouzi, G.~Bruder, and G.~F. Welch, ``Reducing task
  load with an embodied intelligent virtual assistant for improved performance
  in collaborative decision making,'' in \emph{IEEE Conference on Virtual
  Reality and 3D User Interfaces (VR)}, 2020, pp. 529--538.

\bibitem{yao2022virtual}
H.~Yao, A.~G. de~Siqueira, A.~Bafna, D.~Peterkin, J.~Richards, M.~L. Rogers,
  A.~Foster, I.~Galynker, and B.~Lok, ``A virtual human interaction using
  scaffolded ping-pong feedback for healthcare learners to practice empathy
  skills,'' in \emph{Proceedings of the ACM International Conference on
  Intelligent Virtual Agents}, 2022, pp. 1--8.

\bibitem{faul2007g}
F.~Faul, E.~Erdfelder, A.-G. Lang, and A.~Buchner, ``G* power 3: A flexible
  statistical power analysis program for the social, behavioral, and biomedical
  sciences,'' \emph{Behavior research methods}, vol.~39, no.~2, pp. 175--191,
  2007.

\bibitem{kenngott2015openhelp}
H.~Kenngott, J.~W{\"u}nscher, M.~Wagner, A.~Preukschas, A.~Wekerle, P.~Neher,
  S.~Suwelack, S.~Speidel, F.~Nickel, D.~Oladokun \emph{et~al.}, ``Openhelp
  (heidelberg laparoscopy phantom): development of an open-source surgical
  evaluation and training tool,'' \emph{Surgical endoscopy}, vol.~29, no.~11,
  pp. 3338--3347, 2015.

\bibitem{feick2020virtual}
M.~Feick, N.~Kleer, A.~Tang, and A.~Kr{\"u}ger, ``The virtual reality
  questionnaire toolkit,'' in \emph{Adjunct Proceedings of the 33rd Annual ACM
  Symposium on User Interface Software and Technology}, 2020, pp. 68--69.

\bibitem{Brooke1995SUS}
J.~Brooke, ``{SUS}: A quick and dirty usability scale,'' \emph{Usability
  evaluation in industry}, vol. 189, 1995.

\bibitem{bangor2009determining}
A.~Bangor, P.~Kortum, and J.~Miller, ``Determining what individual {SUS} scores
  mean: Adding an adjective rating scale,'' \emph{Journal of usability
  studies}, vol.~4, no.~3, pp. 114--123, 2009.

\bibitem{hart2006nasa}
S.~G. Hart, ``{NASA-Task Load Index (NASA-TLX)}; 20 years later,'' in
  \emph{Proceedings of the human factors and ergonomics society annual
  meeting}, vol.~50, 2006, pp. 904--908.

\bibitem{schubert2001experience}
T.~Schubert, F.~Friedmann, and H.~Regenbrecht, ``The experience of presence:
  Factor analytic insights,'' \emph{Presence: Teleoperators \& Virtual
  Environments}, vol.~10, no.~3, pp. 266--281, 2001.

\bibitem{schwind2019using}
V.~Schwind, P.~Knierim, N.~Haas, and N.~Henze, ``Using presence questionnaires
  in virtual reality,'' in \emph{Proceedings of the 2019 CHI conference on
  human factors in computing systems}, 2019, pp. 1--12.

\bibitem{alkaissi2023artificial}
H.~Alkaissi and S.~I. McFarlane, ``Artificial hallucinations in chatgpt:
  implications in scientific writing,'' \emph{Cureus}, vol.~15, no.~2, 2023.

\bibitem{li2023mmasd}
J.~Li, V.~Chheang, P.~Kullu, E.~Brignac, Z.~Guo, A.~Bhat, K.~E. Barner, and
  R.~L. Barmaki, ``Mmasd: A multimodal dataset for autism intervention
  analysis,'' pp. 397--405, 2023.

\bibitem{hartholt2019ubiquitous}
A.~Hartholt, E.~Fast, A.~Reilly, W.~Whitcup, M.~Liewer, and S.~Mozgai,
  ``Ubiquitous virtual humans: A multi-platform framework for embodied ai
  agents in xr,'' in \emph{IEEE International Conference on Artificial
  Intelligence and Virtual Reality (AIVR)}.\hskip 1em plus 0.5em minus
  0.4em\relax IEEE, 2019, pp. 308--3084.

\bibitem{canal2022survey}
F.~Z. Canal, T.~R. M{\"u}ller, J.~C. Matias, G.~G. Scotton, A.~R. de~Sa~Junior,
  E.~Pozzebon, and A.~C. Sobieranski, ``A survey on facial emotion recognition
  techniques: A state-of-the-art literature review,'' \emph{Information
  Sciences}, vol. 582, pp. 593--617, 2022.

\bibitem{li21twostage}
J.~Li, A.~Bhat, and R.~Barmaki, ``A two-stage multi-modal affect analysis
  framework for children with autism spectrum disorder,'' in \emph{Proceedings
  of the AAAI-21 Workshop on Affective Content Analysis}, 2021, pp. 1--8.

\bibitem{guo2023social}
Z.~Guo, V.~Chheang, J.~Li, K.~E. Barner, A.~Bhat, and R.~Barmaki, ``Social
  visual behavior analytics for autism therapy of children based on automated
  mutual gaze detection,'' in \emph{Proceedings of the International Conference
  on Cooperative and Human Aspects of Software Engineering}, ser. CHASE '23,
  2023.

\bibitem{barmaki2018gesturing}
R.~Barmaki and C.~Hughes, ``Gesturing and embodiment in teaching: Investigating
  the nonverbal behavior of teachers in a virtual rehearsal environment,'' in
  \emph{Proceedings of the AAAI Conference on Artificial Intelligence},
  vol.~32, no.~1, 2018.

\bibitem{li2022pose}
J.~Li, A.~Bhat, and R.~Barmaki, ``Pose uncertainty aware movement synchrony
  estimation via spatial-temporal graph transformer,'' in \emph{Proceedings of
  the International Conference on Multimodal Interaction}, ser. ICMI '22, 2022,
  p. 73–82.

\bibitem{scavarelli2021virtual}
A.~Scavarelli, A.~Arya, and R.~J. Teather, ``Virtual reality and augmented
  reality in social learning spaces: a literature review,'' \emph{Virtual
  Reality}, vol.~25, pp. 257--277, 2021.

\bibitem{chheang2022towards}
V.~Chheang, D.~Schott, P.~Saalfeld, L.~Vradelis, T.~Huber, F.~Huettl, H.~Lang,
  B.~Preim, and C.~Hansen, ``Towards virtual teaching hospitals for advanced
  surgical training,'' in \emph{IEEE Conference on Virtual Reality and 3D User
  Interfaces Abstracts and Workshops (VRW)}.\hskip 1em plus 0.5em minus
  0.4em\relax IEEE, 2022, pp. 410--414.

\end{thebibliography}

\end{document}